\documentclass[12pt]{iopart}

\usepackage[]{cite}

\usepackage[]{amssymb}
\usepackage[]{mathrsfs}
\usepackage[]{eucal}
\usepackage[]{tensor}
\usepackage[]{amsthm}
\usepackage[]{bm}

\newcommand{\dd}{\partial}
\newcommand{\df}{\mathrm{d}}
\newcommand{\w}{\wedge}
\newcommand{\Lie}{\pounds}

\newcommand{\qqd}{\ , \quad}
\newcommand{\hor}{_{\mathsf{H}}}
\newcommand{\bc}{\begin{center}}
\newcommand{\ec}{\end{center}}
\newcommand{\be}{\begin{equation}}
\newcommand{\ee}{\end{equation}}

\newcommand{\FF}{\mathcal{F}}
\newcommand{\GG}{\mathcal{G}}
\newcommand{\LL}{\mathscr{L}}

\usepackage{dsfont}
\newcommand{\nn}{\mathds{N}}

\newcommand{\rr}{\mathds{R}}

\usepackage[unicode]{hyperref}
\usepackage{xcolor}
\definecolor{pastgreen}{HTML}{669900}
\definecolor{pastblue}{HTML}{336699}
\definecolor{linkcol}{HTML}{663333}
\hypersetup{colorlinks,linkcolor={pastblue},citecolor={pastblue},urlcolor={pastblue}}

\theoremstyle{plain} \newtheorem{tm}{Theorem}[section]
\theoremstyle{plain} \newtheorem{lm}[tm]{Lemma}
\theoremstyle{plain} \newtheorem{cor}[tm]{Corollary}
\theoremstyle{definition} \newtheorem{defn}[tm]{Definition}
\theoremstyle{definition} 
\newcommand{\btm}{\begin{tm}}
\newcommand{\etm}{\end{tm}}
\newcommand{\blm}{\begin{lm}}
\newcommand{\elm}{\end{lm}}
\newcommand{\bcor}{\begin{cor}}
\newcommand{\ecor}{\end{cor}}
\newcommand{\bdefn}{\begin{defn}}
\newcommand{\edefn}{\end{defn}}

\begin{document}

\begin{flushright}
ZTF-EP-17-09
\end{flushright}

\title[Generalizations of the Smarr's formula \dots]{Generalizations of the Smarr formula for black holes with nonlinear electromagnetic fields}

\author{Luka Gulin and Ivica Smoli\'c}
\address{Department of Physics, Faculty of Science, University of Zagreb, Bijeni\v cka cesta 32, 10000 Zagreb, Croatia}
\eads{\mailto{lgulin@phy.hr}, \mailto{ismolic@phy.hr}}

\date{\today}

\begin{abstract}
We present a direct, geometric derivation of the generalized Smarr formula for the stationary axially symmetric black holes with nonlinear electromagnetic fields. The additional term is proven to be proportional to the integral of the trace of the electromagnetic energy-momentum tensor and can be written as a product of two conjugate variables. From the novel relation we can deduce all previously proposed forms of the generalized Smarr formula, which were derived only for the spherically symmetric black holes, and provide the lowest order quantum correction to the classical relation from the Euler-Heisenberg Lagrangian.
\end{abstract}


\vspace{2pc}



\section{Introduction}

The classical thermodynamic Euler relation \cite{Callen,Greiner,Sethna} or, as some authors call it \cite{Schwabl}, the Gibbs-Duhem equation,
\be\label{eq:GD}
E = TS + x_i X^i \ ,
\ee
provides a constraint between the energy $E$, the temperature $T$, the entropy $S$ and the rest of the pairs $\{(x_i, X^i)\}$ of the conjugate, intensive and extensive, thermodynamic quantities that describe the system. The usual proof of this relation rests upon the assumption that the energy $E = E(S;\{X^i\})$ is a smooth homogeneous function of \emph{degree 1} and the Euler's homogeneous function theorem \cite{Lee}, which states that $k f(\mathbf{X}) = \mathbf{X}\cdot\bm{\nabla}f(\mathbf{X})$ holds for any smooth homogeneous function $f : (\rr^n)^\times \to \rr$ of degree $k$. Then we have 
\be
E = \frac{\dd E}{\dd S}\,S + \frac{\dd E}{\dd X^i}\,X^i \ ,
\ee
and the partial derivatives can be extracted from the first law of thermodynamics, resulting in (\ref{eq:GD}). Amazingly, as was first noted by Smarr \cite{Smarr72}, the analogous formula holds for the Kerr-Newman black hole,
\be\label{eq:KNSmarr}
M = 2T\hor S + 2\Omega\hor J + \Phi\hor Q\hor \ ,
\ee
relating the mass $M$, the temperature $T\hor = \kappa/2\pi$ (where $\kappa$ is the surface gravity), the entropy $S = \mathcal{A}\hor/4$ (where $\mathcal{A}\hor$ is the area of the black hole horizon), the angular velocity of the horizon $\Omega\hor$, the angular momentum $J$, the electrostatic potential at the horizon $\Phi\hor$ (gauged so that it is zero at the infinity), and the charge $Q\hor$. The origin of the factor $2$ in front of the $T\hor S$ and the $\Omega\hor J$ terms comes from the fact that the mass of the Kerr-Newman black hole, written in the form $M = M(A\hor,J,Q\hor^2)$, is a homogeneous function of \emph{degree} $\mathit{1/2}$ (see also comments in \cite{Natsuume}, Appendix to Chapter 3). This discrepancy with respect to the classical result should not come as a surprise once we are aware that we are looking at the system in the presence of the long-range, gravitational interactions.

\medskip

The first step of generalization of the original Smarr's result was provided by the geometric\footnote{We use adjective ``geometric'' just to emphasize that in these derivations auxiliary physical assumptions are avoided by the approach based on the quantities of geometric origin.} derivation \cite{C73,C87,Heusler97,Heusler}, based on the Komar integrals \cite{Komar63}. This has demonstrated that the relation (\ref{eq:KNSmarr}) does not depend on the particular details of the Kerr-Newman black hole (see also \cite{CG17} for a recent discussion about the dyonic Kerr-Newman black hole). Namely, the result is independent of the Eulerian scaling argument (we do not have to assume \emph{a priori} that $M$ is a homogeneous function), and independent of the first law of the black hole mechanics (which may bring in the procedure some subtle issues \cite{MZ14}). The Smarr formula thus stands among the pillars of the black hole thermodynamics, an object of intensive research for the past four decades \cite{Wald99} which reveals a deep relationship between the gravitation, thermodynamics and quantum theory.

\medskip

Apart from the black holes that appear in the solutions of the Einstein-Max\-well's field equations, there are also black hole solutions in theories with nonlinear electrodynamics (NLE). These models were introduced almost a century ago by Born and Infeld \cite{Born34,BI34}, in a quest for the alternative to the Maxwell's electrodynamics which could cure its inconsistencies associated with the infinite self-energy of the point charges, but it was soon realized that such nonlinearities appear in the quantum corrections to classical electromagnetic interaction \cite{HE36}. This type of models have been also used to regularize the black hole \cite{Bardeen68,ABG98,ABG00,Bronnikov00,BH02} and the cosmological singularities \cite{GSB00,GSB04,CGMCL04}, as well as to mimic the cosmological ``dark energy'' \cite{ELNO03,LR10,BLM12}. The nonlinearities in the electromagnetic fields will be probed in the numerous forthcoming experiments \cite{BR13,FBR16,MCLP17,EMY17}.

\medskip

The earliest attempt of the generalization of the Smarr formula to nonlinear models of electrodynamics was presented by Rasheed in \cite{Rasheed97}. However, the result was inconclusive, since after the initial sketch the author just states that the Smarr formula in the presence of the NLE fields cannot hold any more in its original simple form, thus leaving much to be desired. All subsequent analyses \cite{Breton04,GHM09,YH10,GMK12,DARG13,ZG16,FW16} of the thermodynamic properties of black holes with NLE approach the generalization of the Smarr formula for some special subcases, which were basically just the spherically symmetric black holes, and this was always done via Eulerian scaling argument.

\medskip

The main objective of this paper is to derive the generalized Smarr formula, valid for any stationary axially symmetric black hole with the electromagnetic field defined by some of the nonlinear models. By utilizing the geometric approach we shall avoid pitfalls and limitations of the previous efforts, which will allow us to give a bird's-eye view on this aspect of the black hole electrodynamics and thermodynamics. One of our aims is to gain a better understanding of the conditions under which the relations of the Smarr type hold at all.

\medskip

Before doing any concrete calculation we need do clear out what is the sought form of the result. We shall say that the theory admits a generalization of the Smarr formula if the charges are related by a formula of the form
\be
M = \ell(T\hor S, \Omega\hor J, \Phi\hor Q\hor, \dots) + h(\Phi\hor Q\hor, \dots) \ ,
\ee
which consists of the linear part, 
\be
\ell = c_1 T\hor S + c_2 \Omega\hor J + c_3 \Phi\hor Q\hor + \dots \ ,
\ee
defined by a set of real constants $\{c_1,c_2,\dots\}$, and a possible additional nonlinear function $h$ which depends solely on the ``horizon data'' (i.e.~products of various pairs of intensive and extensive variables evaluated at the black hole horizon). However, a typical interim result will be of the form $h = h(M,\Phi\hor Q\hor,\dots)$, leaving relation between the mass $M$ and the rest of the variables in an implicit form (possibly with highly nontrivial corresponding explicit relation).

\medskip

The paper is organized as follows. In section 2 we explain in detail all the basic assumptions that will be used throughout the paper. Section 3 is the central part of the paper with the derivation of the generalized Smarr formula for the nonlinear electromagnetic models. In section 4 we rewrite the additional term on the generalized Smarr formula as a product of a conjugate pair of an extrinsic and an intrinsic variable and in section 5 we use our relation to generalize the Smarr formula for the stationary axisymmetrc black holes with power-Maxwell electrodynamics. In section 6 we overview the spherically symmetric cases and derive the lowest order quantum correction to the Smarr formula coming from the Euler-Heisenberg Lagrangian. In the final section we summarize the results and discuss the most important open question. Finally, in the appendices we collect basic facts about the NLE models and some useful formulae from the differential geometry that were used throughout the paper.

\medskip

\emph{Notation, conventions, remarks}. We shall use the metric signature $(-,+,+,+)$ and the natural systems of units with $G = c = 4\pi\varepsilon_0 = 1$. We use the abstract index notation (as in \cite{Wald}) whenever the type of the tensor in an equation has to be emphasized, and the ``indexless'' notation (as in \cite{Heusler}) to simplify most of the calculations (only some final examples are done in concrete coordinate systems). For the contraction of a symmetric tensor $S_{ab}$ with a vector $X^a$ we use the notation $S(X)_a \equiv S_{ab} X^b$. Electromagnetic field 2-form $F_{ab}$ may be locally introduced via gauge 1-form $A_a$ as $F = \df A$, while its Hodge dual ${*F}_{ab}$ is given by
\be
{*F}_{ab} = \frac{1}{2}\,F_{cd}\,\tensor{\epsilon}{^c^d_a_b} \ .
\ee
The two electromagnetic invariants are denoted by
\be
\FF \equiv F_{ab} F^{ab} \qquad \textrm{and} \qquad \GG \equiv F_{ab}\,{*F}^{ab} \ .
\ee
We are looking at the class of nonlinear electromagnetic (NLE) models described by the Lagrangian of the form
\be
L = \LL(\FF,\GG)\,{*1} \ .
\ee
For example, the canonical Maxwell's electromagnetic Lagrangian density is $\LL^{\mathrm{(Max)}} = -\FF/4$. We denote the partial derivatives with
\be
\LL_\FF \equiv \frac{\dd\LL}{\dd\FF} \qqd \LL_\GG \equiv \frac{\dd\LL}{\dd\GG} \qqd \textrm{etc.}
\ee
By abuse of language we will usually refer to $\LL$ as a ``Lagrangian'' instead of Lagrangian density.

\bigskip

\section{Basic elements of the analysis}

Let us first lay out the geometric setting upon which we will derive the Smarr formula. We are looking at the smooth, asymptotically flat 4-dimensional spacetime $(M,g_{ab},F_{ab})$, which is a solution of the source-free Einstein-(generalized) Maxwell field equations,
\be
G_{ab} = 8\pi T_{ab} \ ,
\ee
\be\label{eq:gMax}
\df F = 0 \ , \quad \textrm{and} \qquad \df\,{*Z} = 0 \ ,
\ee
where the energy-momentum tensor is
\be\label{eq:NLT}
T_{ab} = -\frac{1}{4\pi} \Big( (\LL_\GG \GG - \LL) g_{ab} + 4 \LL_\FF \, F_{ac} \tensor{F}{_b^c} \Big) \ ,
\ee
and the auxiliary 2-form $Z_{ab}$ is defined as\footnote{For a general normalization $\LL^{\mathrm{(Max)}} = \mu\FF$, with some real constant $\mu\ne 0$, we would choose $Z_{ab} \equiv \mu^{-1} \left( \LL_\FF\,F_{ab} + \LL_\GG\,{*F}_{ab} \right)$.}
\be
Z_{ab} \equiv -4 \left( \LL_\FF\,F_{ab} + \LL_\GG\,{*F}_{ab} \right) \ .
\ee
Factors $\LL_\FF$ and $\LL_\GG$ are the source of the nonlinearity in the second generalized Maxwell's equation (\ref{eq:gMax}). A convenient way to rewrite the energy-momentum tensor (\ref{eq:NLT}),
\be\label{eq:NLTMax}
T_{ab} = -4\LL_\FF T^{\mathrm{(Max)}}_{ab} + \frac{1}{4}\,T g_{ab} \ ,
\ee
is done with help of the canonical Maxwell's energy-momentum tensor,
\be\label{eq:TMax}
T^{\mathrm{(Max)}}_{ab} = \frac{1}{4\pi} \left( F_{ac}\tensor{F}{_b^c} - \frac{1}{4}\,g_{ab}\FF \right)
\ee
and the trace of the energy-momentum tensor,
\be\label{eq:emTr}
T \equiv g^{ab} T_{ab} = \frac{1}{\pi} \left( \LL - \LL_\FF \FF - \LL_\GG \GG \right) \ .
\ee
Note that in the Maxwell's case, $\LL^{\mathrm{(Max)}} = -\FF/4$, we have $\LL_\GG = 0$ and $\LL_\FF \FF = \LL$, thus the energy-momentum tensor (\ref{eq:TMax}) is traceless.

\medskip

The spacetime is assumed to be stationary axisymmetric with the corres\-pond\-ing stationary Killing vector $k^a$ and the axial Killing vector $m^a$. Without any loss of generality we can assume that these two Killing vector fields commute \cite{Carter70,Szabados87}. If the spacetime is axially symmetric we assume that the axis of symmetry (the set where $m^a = 0$) is nonempty and intersects the domain of outer communications. Furthermore, we assume that the electromagnetic field inherits all symmetries,
\be\label{eq:syminhF}
\Lie_\xi F_{ab} = 0
\ee
for any Killing vector field $\xi^a$. This, seemingly natural, is a highly nontrivial assumption which was analysed for the canonical Maxwell's electrodynamic in \cite{MW75,Tod06} and more recently for the nonlinear electromagnetic fields in \cite{BGS17}. Given any Killing vector field $\xi^a$ we can introduce the electric 1-form $E = -i_\xi F$ and the magnetic 1-form $H = i_\xi {*Z}$. Due to the generalized Maxwell's equation (\ref{eq:gMax}) and the assumption about the symmetry inheritance (\ref{eq:syminhF}), these are closed forms, $\df E = 0 = \df H$. The Poincar\'e lemma implies that at least \emph{locally} we can introduce the electric scalar potential, via $E = -\df\Phi$, and the magnetic scalar potential, via $H = -\df\Psi$. In order to guarantee that the electromagnetic scalar potentials are \emph{globally} well-defined we shall assume that the domain of outer communications is simply connected, so that its first de Rham cohmology group is trivial (see e.g.~theorem 15.17 in \cite{Lee}). We always make a gauge choice such that both $\Phi$ and $\Psi$ vanish at the infinity. Note that there is also a sign ambiguity in the definition of the scalar potentials: whereas we are following here the usual, traditional choice, part of the literature (e.g.~\cite{Heusler}) contains the opposite choice (and compensate this with some other unconventional sign choices in order to get the Smarr formula in its standard form). 

\medskip

Finally, the spacetime contains a connected Killing horizon $H[\chi]$, generated by the Killing vector field $\chi^a = k^a + \Omega\hor m^a$, where $\Omega\hor$ is the ``angular velocity'' of the horizon \cite{C73,Heusler} (and $\Omega\hor = 0$ in the case of nonrotating Killing horizon $H[k]$). All the proofs of the Smarr formula are based upon several crucial assumptions about the equipotential nature of the black hole horizons, which are supported by the following well-known theorems:

\begin{itemize}
\item[(a)] the zeroth law of black hole thermodynamics (the surface gravity $\kappa$ is constant on each component the Killing horizon \cite{KayWald91,RW95,Rasheed97}),
\item[(b)] the weak rigidity (the angular velocity of each component of the Killing horizon is constant \cite{C73,Heusler,HCC}),
\item[(c)] the zeroth law of black hole electrodynamics (electromagnetic scalar potentials are constant on each component of the Killing horizon \cite{C73,Rasheed97,ISm12,ISm14,BGS17}).
\end{itemize}

As this will turn out to be the most convenient choice, we choose to define the electric and the magnetic field, as well as the corresponding scalar potentials $\Phi$ and $\Psi$, with respect to the Killing vector field $\chi^a$. Since $\Omega\hor$ is a constant, the vector field $\chi^a$ is well-defined on the same domain as $k^a$ and $m^a$.

\bigskip

\section{Geometric approach to the generalized Smarr formula}

The basic quantities related by the Smarr formula may be introduced via well-known Komar integrals \cite{Heusler,JG11}. For any smooth closed 2-surface $\mathcal{S}$ we define, respectively, the mass $M_\mathcal{S}$, the angular momentum $J_\mathcal{S}$, the electric charge $Q_\mathcal{S}$ and the magnetic charge $P_\mathcal{S}$ via
\be\label{eq:KomarMJS}
M_{\mathcal{S}} = -\frac{1}{8\pi} \int_{\mathcal{S}} {*\df k} \qqd J_{\mathcal{S}} = \frac{1}{16\pi} \int_{\mathcal{S}} {*\df m} \ ,
\ee
\be\label{eq:KomarQP}
Q_{\mathcal{S}} = \frac{1}{4\pi} \int_{\mathcal{S}} {*Z} \ , \quad \textrm{and} \qquad P_{\mathcal{S}} = \frac{1}{4\pi} \int_{\mathcal{S}} F \ .
\ee
Let $\Sigma \subseteq M$ be a smooth spacelike hypersurface, extending from the spacelike infinity to the horizon $H[\chi]$, which it intersects in the closed 2-surface $\mathcal{H} = \Sigma \cap H[\chi]$. The two most important choices of the surface $\mathcal{S}$ will be $\mathcal{S} = \mathcal{H}$ (in which case we shall denote the corresponding Komar quantities by $M\hor$, $J\hor$, $Q\hor$ and $P\hor$) and $\mathcal{S} = \mathbb{S}^2_\infty$, the 2-sphere at the spacelike infinity (in which case the corresponding Komar quantities will be denoted by a letter without any index, $M$, $J$, $Q$ and $P$). Note, however, that $Q = Q\hor$ and $P = P\hor$, as a direct consequence of the Stokes' theorem and the fact that the generalized Maxwell's equations (\ref{eq:gMax}) are source-free. It can be shown \cite{Heusler} that the Komar integrals (\ref{eq:KomarMJS}) for these two choices of the integration surface, can be written in the following way,
\be
M - M\hor = -\frac{1}{4\pi} \int_\Sigma {*R}(k) \ ,
\ee
\be
J - J\hor = \frac{1}{8\pi} \int_\Sigma {*R}(m) \ .
\ee
Then, using the Einstein's gravitational field equation
\be
R_{ab} = 8\pi \left( T_{ab} - \frac{1}{2}\,T g_{ab} \right) \ ,
\ee
we get the formula
\be
M - M\hor - 2\Omega\hor(J - J\hor) = -2 \int_\Sigma \left( {*T}(\chi) - \frac{1}{2}\,T\,{*\chi} \right) \ .
\ee
Furthermore, there is also an additional geometric relation, closely related to the Smarr formula (\ref{eq:KNSmarr})
\be
M\hor = \frac{\kappa \mathcal{A}\hor}{4\pi} + 2\Omega\hor J\hor \ .
\ee
By combining these two equations we can express the total mass $M$ as
\be\label{eq:genSmarr}
M = \frac{\kappa \mathcal{A}\hor}{4\pi} + 2\Omega\hor J - 2 \int_\Sigma \left( {*T}(\chi) - \frac{1}{2}\,T\,{*\chi} \right) \ .
\ee
This formula, which in slightly varied forms is sometimes referred to as a Bardeen-Carter-Hawking mass formula \cite{BCH73,HS93,Visser93}, is a major intermediate step in the derivation of the Smarr relation in the presence of any matter or gauge field described by the energy-momentum tensor $T_{ab}$. 

\medskip

With help of the auxiliary identity (\ref{eq:aux1}) we can write the Maxwell's electromagnetic energy-momentum tensor (\ref{eq:TMax}) in a following form,
\be\label{eq:TMaxFFFF}
8\pi T^{\mathrm{(Max)}}_{ab} = F_{ac}\tensor{F}{_b^c} + {*F}_{ac} \tensor{{*F}}{_b^c} \ .
\ee
We shall use here the electric 1-form $E_a$ and the magnetic 1-form $H_a$, introduced via
\be
E = -i_\chi F \qquad \textrm{and} \qquad H = i_\chi\,{*Z} \ .
\ee
Let us look at the following 1-form
\be
*(E \w {*Z} + H \w F) = -i_E Z + i_H {*F} \ .
\ee
On one hand we have
\be
E \w {*Z} + H \w F = -\df (\Phi\,{*Z} + \Psi F) \ .
\ee
On the other hand, using (\ref{eq:TMaxFFFF}),
\be
*(E \w {*Z} + H \w F)_a = 32\pi\,\LL_\FF\,T^{\mathrm{(Max)}}(\chi)_a \ .
\ee
Thus, using (\ref{eq:NLTMax}), we can write
\be
8\pi \left( T(\chi)_a - \frac{1}{2}\,T \chi_a \right) = {*\df}\,(\Phi\,{*Z} + \Psi F)_a - 2\pi T \chi_a \ .
\ee
So, if we assume some appropriate fall-off conditions, for example that $\Phi\,{*Z} = O(r^{-\epsilon})$ and $\Psi F = O(r^{-\epsilon})$ for some real $\epsilon > 0$ (so that the term at the infinity drops), it follows that
$$-\frac{1}{4\pi} \int_\Sigma \df (\Phi\,{*Z} + \Psi F) = -\frac{1}{4\pi} \left( \int_{\mathbb{S}^2_\infty} - \int_{\mathcal{H}} \right) (\Phi\,{*Z} + \Psi F) =$$
\be\label{eq:QP}
= \frac{1}{4\pi} \int_{\mathcal{H}} (\Phi\,{*Z} + \Psi F) = \Phi\hor Q\hor + \Psi\hor P\hor \ ,
\ee
where we have used the fact that $\Phi\hor$ and $\Psi\hor$ are constant over $H[\chi]$. Inserting this into the formula (\ref{eq:genSmarr}) we get the generalized Smarr formula
\be\label{eq:nleSmarr}
M = \frac{\kappa \mathcal{A}\hor}{4\pi} + 2\Omega\hor J + \Phi\hor Q\hor + \Psi\hor P\hor + \Delta \ ,
\ee
where
\be\label{eq:Delta}
\Delta = \frac{1}{2} \, \int_\Sigma T\,{*\chi} \ .
\ee
This is the central result of the paper. Furthermore, since by physical prescription in the black hole thermodynamics $T\hor = \kappa/2\pi$ and $S = \mathcal{A}\hor/4$, the equation (\ref{eq:nleSmarr}) can be equivalently written as
\be
M = 2T\hor S + 2\Omega\hor J + \Phi\hor Q\hor + \Psi\hor P\hor + \Delta \ .
\ee

If the spacetime is static with the nonrotating Killing horizon $H[k]$, we can repeat verbatim the complete proof from above, just by inserting $\Omega\hor = 0$, thereby reducing $\chi^a$ to the Killing vector field $k^a$. Since here we don't need the notion of the angular momentum $J$, the resulting generalized Smarr formula for the static black holes remains valid even if the spacetime is not necessarily axially symmetric. Note, however, that in the static non-axially symmetric case we rely on the field equation proof \cite{Rasheed97,ISm12,ISm14} of the zeroth law of the electrodynamics. Furthermore, if we have a static multi-black hole case, that is the case when the nonrotating Killing horizon $H[k]$ is a union of $N\in\nn$ connected components $H_i[k]$, then the scalar electromagnetic potentials are constant on each connected component $H_i[k]$ (but do not necessarily have to pairwise equal). If we denote each potential and charge of the corresponding component $H_i[k]$ by the index ``$i$'', then we have a decomposition
\be
\int_{\mathcal{H}} (\Phi {*F} + \Psi F) = \sum_{i=1}^N (\Phi_i Q_i + \Psi_i P_i) \ ,
\ee
and on each connected component $H_i[k]$ the mass $M_i$, the surface gravity $\kappa_i$, and the area $\mathcal{A}_i$ of $H_i[\chi] \cap \Sigma$ are related (see e.g.~equation (6) in \cite{Heusler96}) by the equation
\be
M_i = \frac{\kappa_i \mathcal{A}_i}{4\pi} \ .
\ee
Using these two relations we immediately have a generalization of the Smarr formula for the static nonconnected case (compare it with the equation (19) in \cite{Heusler96}),
\be
M = \sum_{i=1}^N \left( \frac{\kappa_i \mathcal{A}_i}{4\pi} + \Phi_i Q_i + \Psi_i P_i \right) + \Delta \ .
\ee

\medskip

Obviously, in the case of the Maxwell's electrodynamics the energy-momentum tensor is traceless and we immediately have $\Delta = 0$. Conversely, if we take the condition $T = 0$ as a partial differential equation for the Lagrangian $\LL(\FF,\GG)$, then on the set where $\FF \ne 0$ the general solution is of the form $\LL = f(\GG/\FF)\FF$, with some differential real function $f$, while on the set where $\GG \ne 0$ the general solution is of the form $\LL = g(\FF/\GG)\GG$, with some differential real function $g$. It is difficult to say if such class of models, apart from the canonical Maxwell's, has any physical merit.

\medskip

In the general nonlinear case $\LL = \LL(\FF,\GG)$, the difficulty of evaluating the integral in (\ref{eq:Delta}) and reducing it to the quantities at the horizon (and possibly those at the infinity) is directly related to the problem of writing the integrand in $\Delta$ as an exact form, which would allow us to use the Stokes' theorem. By the symmetry inheritance, the integrand is a closed form,
\be
*\df (T\,{*\chi}) = *(\df T \w {*\chi}) + T\,{*\df}{*\chi} = -i_\chi\df T = -\Lie_\chi T = 0 \ .
\ee
Hence, due to the Poincar\'e lemma, we know that at least \emph{locally} we can write $T\,{*\chi} = \df\sigma$ for some 2-form $\sigma$. This, however, does not have to necessarily hold \emph{globally}, unless some additional topological assumptions are imposed (see discussion in \cite{Wald90,Torre97}). Whenever the 3-form $T\,{*\chi}$ is indeed an exact form, we can use Stokes' theorem to write $\Delta$ in (\ref{eq:Delta}) as a difference between the term at the infinity and a term at the black hole horizon.

\bigskip

\section{Correction as a product of a conjugate pair}

Since all the terms on the right hand side of the generalized Smarr relation (\ref{eq:nleSmarr}) are written as products of conjugate thermodynamics variables one might wish to bring the additional term $\Delta$ to the same form. Indeed, this has been achieved via Eulerian scaling argument for the static, spherically symmetric black holes: first for the truncated Born-Infeld Lagrangian (\ref{eq:tBI}) in \cite{YH10,GMK12} and later \cite{FW16} for the more general class of Lagrangians $\LL = \LL(\alpha,\FF)$ with a physical parameter $\alpha$. Note, however, that all these results rely upon the assumption about the specific form of the first law of black hole mechanics which should hold in the presence of NLE, a proof of which has appeared in a recent unpublished paper \cite{ZG16}. According to the argument presented in \cite{FW16} the additional term in the Smarr formula should be proportional to the product of the parameter $\alpha$ and an auxiliary quantity $\Pi$, defined by a volume integral\footnote{there is probably a typo in their formula (44), which should include an additional integral over the angular variables} of the partial derivative $\dd\LL/\dd\alpha$. We shall now demonstrate how to write $\Delta$, defined by (\ref{eq:Delta}), exactly in this form. 

\medskip

All the examples of the Lagrangians presented in \cite{FW16} are of the form $\LL(\alpha,\FF) = \alpha^{-1} \widetilde{\LL}(\alpha\FF)$, where $\widetilde{\LL}$ is a real differentiable function. In this case a simple relation holds,
\be
\LL - \LL_\FF \FF = -\alpha\,\frac{\dd\LL}{\dd\alpha} \ ,
\ee
so that in turn we immediately have
\be\label{eq:DeltaalphaPi}
\Delta = -\frac{\alpha}{2\pi}\,\int_\Sigma \frac{\dd\LL}{\dd\alpha}\,{*\chi} \ ,
\ee
which proves the claim. Again, our approach is independent of the first law and immediately valid for the stationary axially symmetric spacetimes, such are those with the rotating black holes. In fact, we can take this relation one step further towards more general relation. Suppose that we have a class of Lagrangians of the form $\LL(\beta,\FF,\GG) = \beta^{-1} \widetilde{\LL}(\beta\FF,\beta\GG)$, where $\beta$ is a physical parameter and $\widetilde{\LL}$ is a real differentiable function of two variables. Here we have
\be\label{eq:LFLGbeta}
\LL - \LL_\FF \FF - \LL_\GG \GG = -\beta\,\frac{\dd\LL}{\dd\beta} \ ,
\ee
so that
\be
\Delta = \beta \mathcal{C}
\ee
holds, where we have introduced another auxiliary quantity
\be
\mathcal{C} \equiv -\frac{1}{2\pi}\,\int_\Sigma \frac{\dd\LL}{\dd\beta}\,{*\chi} \ .
\ee
For example, if we write Euler-Heisenberg Lagrangian (\ref{eq:EH}) in the form
\be
\LL = \frac{1}{\gamma} \left( -\frac{1}{4}\,\gamma\FF + \left( 4(\gamma\FF)^2 + 7(\gamma\GG)^2 \right) \right)
\ee
we get
\be\label{eq:EHDeltaC}
\Delta_{\mathrm{EH}} = -\frac{\gamma}{2\pi} \int_\Sigma \left( 4\FF^2 + 7\GG^2 \right)\,{*\chi} \ .
\ee
This is the lowest order contribution to the Smarr formula coming from the quantum electrodynamic effects.

\medskip

The crucial question here is: What is the physical interpretation of the conjugate pair $(\beta,\mathcal{C})$? First, equations (\ref{eq:Delta}) and (\ref{eq:LFLGbeta}) show that $\mathcal{C}$ is proportional to the integral of trace $T$ of the energy-momentum tensor. From this perspective, it seems that $\mathcal{C}$ may play the similar role as the cosmological constant in the black hole thermodynamics \cite{ELNO03,LR10,KMT17}. Second, note that there is an ambiguity in the choice of the parameter $\beta$. For example, if we use an alternative parametrization $\beta = b^\lambda$, with some new physical parameter $b$ and $\lambda\in\rr^\times$, we will have
\be
\Delta = b\mathcal{C}_\lambda \qquad \mathrm{with} \qquad \mathcal{C}_\lambda \equiv -\frac{1}{2\pi\lambda}\,\int_\Sigma \frac{\dd\LL}{\dd b}\,{*\chi} \ .
\ee
As it was noticed in \cite{GMK12}, $\mathcal{C}_{-2}$ for the Born-Infeld NLE (\ref{eq:BI}) has dimensions of the polarization density, so it is tempting to interpret it as some form of the NLE (vacuum) polarization.

\bigskip

\section{The power-Maxwell electromagnetic Lagrangian}

The $\Delta$ term for the power-Maxwell model (\ref{eq:pM}) can be relatively easily directly evaluated in the generality of the geometric setting that we are pursuing here. Suppose first that the Lagrangian is of the form $\LL = \LL(\FF)$. Then, using (\ref{eq:alphabeta}),
\be
F \w {*Z} = -2\LL_\FF\,\FF \, {*1} \ .
\ee
If we contract the last equation with $\chi^a$, using (\ref{eq:iXHodge}) we get
\be
\df (\Phi\,{*Z} - \Psi F) = -2\LL_\FF\,\FF \, {*\chi} \ .
\ee
More concretely, for $\LL = C \FF^s$ we have $\LL_\FF \FF = sC\FF^s$ and $\LL - \LL_\FF \FF = (1-s)C\FF^s$, thus
\be
\Delta_{\mathrm{pM}} = \frac{(1-s)C}{2\pi} \int_\Sigma \FF^s\,{*\chi} = -\frac{1-s}{4\pi s}\, \int_\Sigma \df (\Phi\,{*Z} - \Psi F) \ ,
\ee
and after the integration as in (\ref{eq:QP}) we get
\be
\Delta_{\mathrm{pM}} = \frac{1-s}{s}\,(\Phi\hor Q\hor - \Psi\hor P\hor) \ .
\ee
In other words, the power-Maxwell class of models allows linear generalization of the Smarr formula (such that the nonlinear part $h$ is exactly zero),
\be
M = \frac{\kappa \mathcal{A}\hor}{4\pi} + 2\Omega\hor J + \frac{1}{s}\,\Phi\hor Q\hor + \left( 2 - \frac{1}{s} \right) \Psi\hor P\hor \ .
\ee
This result is consistent with the generalized Smarr relation obtained in \cite{GHM09,DARG13} for the electrically charged, spherically symmetric black holes. Here we must emphasize that the convergence of the integrals used above implicitly imposes some constraints on the parameter $s$ for which the electromagnetic field is ``well-behaved'' at the infinity (e.g.~in the spherically symmetric case we must have $s \in \left< 1/2, 3/2 \right>$, see next section). As expected, with the $s = 1$ choice we are back again at the canonical Smarr formula.

\bigskip

\section{Spherically symmetric black holes}

Let us turn to the special case of the static, spherically symmetric spacetimes. Most of the known static black hole solutions with the NLE \cite{deO94,ABG98,ABG99,ABG00,YT00,FK03,Dey04,DARG09,RWX13,FW16} have the spacetime metric of the form
\be
\df s^2 = -f(r)\,\df t^2 + \frac{\df r^2}{f(r)} + r^2 (\df\theta^2 + \sin^2\theta\,\df\varphi^2) \ ,
\ee
written in the usual spherical coordinate system \cite{Wald}. As above, we denote the stationary Killing vector with $k^a = (\dd/\dd t)^a$.

\bigskip

\subsection{Several general remarks}

Let us look at the details of the electromagnetic field in this spacetime. First, the theorem 5.6 from \cite{Heusler}, generalized for the NLE in\cite{BGS17}, implies that $F(k,X_{(i)}) = 0$ and ${*F}(k,X_{(i)}) = 0$ for any of the three Killing vector fields $X^a_{(i)}$ that generate the $SO(3)$ isometry. The electric 1-form $E = -i_k F$ and the magnetic 1-form $B = i_k\,{*F}$ (more convenient at this point than the 1-form $H = i_k\,{*Z}$), we can write the electromagnetic 2-form as
\be
F = -E_r(r)\,\df t \w \df r + B_r(r) r^2\sin\theta\,\df\theta \w \df\varphi \ ,
\ee
and the corresponding Hodge dual as
\be
{*F} = E_r(r) r^2\sin\theta\,\df\theta \w \df\varphi + B_r(r)\,\df t \w \df r \ .
\ee
The two electromagnetic invariants are given by
\be
\FF = 2(B_r^2 - E_r^2) \qquad \textrm{and} \qquad \GG = 4 E_r B_r \ .
\ee
The Maxwell's equations (\ref{eq:gMax}) can be written as
\be
\dd_\mu (\sqrt{-g} \, {*F}^{\mu\nu}) = 0 \ ,
\ee
\be
\dd_\mu \Big( \sqrt{-g} \, \left( \LL_\FF\,F^{\mu\nu} + \LL_\GG\,{*F}^{\mu\nu} \right) \Big) = 0 \ .
\ee
In the spherically symmetric case the only nontrivial components are those for $\nu = t$, which can be directly integrated to give
\be\label{eq:sphBE}
B_r = \frac{P}{r^2} \qquad \textrm{and} \qquad \LL_\FF E_r - \LL_\GG B_r = -\frac{Q}{4r^2} \ .
\ee
The integration constants are fixed by the definitions of the charges (\ref{eq:KomarQP}). There are two linearly independent components of the Einstein's equation, $\tensor{G}{^t_t} = 8\pi \tensor{T}{^t_t}$ and $\tensor{G}{^\theta_\theta} = 8\pi \tensor{T}{^\theta_\theta}$, which respectively read
\begin{eqnarray}
(r(f-1))' & = 2r^2 \Big( \LL + 4(\LL_\FF E_r - \LL_\GG B_r) E_r \Big) \ , \label{eq:GTtt} \\
(r^2 f')' & = 4r^2 \Big( \LL - 4(\LL_\FF B_r + \LL_\GG E_r) B_r \Big) \ . \label{eq:GTthth}
\end{eqnarray}
Have we used the slightly more general Ansatz with $g_{tt} = -e^{-2\delta(r)} f(r)$, the difference of the two components of the Einstein's equation, $\tensor{G}{^t_t} - \tensor{G}{^r_r} = 8\pi (\tensor{T}{^t_t} - \tensor{T}{^r_r})$, would again imply that $\delta(r)$ is a constant, which by the asymptotic conditions may be put to zero.

\medskip

Our main interest are the integrals over the hypersurface $\Sigma$ which intersects the horizon. As the original coordinate system $\{t,r,\theta,\varphi\}$ becomes singular at the black hole horizon we need to resort for some other coordinate system, regular at the horizon. The simplest choice is that of the ingoing Eddington-Finkelstein coordinates $\{v,r,\theta,\varphi\}$, with the ingoing null coordinate $v = t + r_*$ and the tortoise coordinate $r_*$, defined by $\df r_* = \df r/f(r)$, so that the spacetime metric becomes
\be
\df s^2 = -f(r)\,\df v^2 + 2\,\df v\,\df r + r^2 (\df\theta^2 + \sin^2\theta\,\df\varphi^2) \ .
\ee
The most important detail is that the Hodge dual of the Killing 1-form $k_a$ in this coordinate system has the form
\be
{*k} = r^2 \sin\theta\,\df r \w \df\theta \w \df\varphi \ .
\ee
By the assumed symmetry inheritance we have $T = T(r)$, hence the formula for $\Delta$ can be now written (using $v = \mathrm{const.}$ hypersurface $\Sigma$) as
\be\label{eq:sphDelta}
\Delta = 2 \int_{r\hor}^\infty (\LL - \LL_\FF \FF - \LL_\GG \GG) r^2\,\df r \ ,
\ee
where $r\hor$ is the radius of the black hole horizon. So, if it is possible to find a primitive function
\be
\tau(r) = \int^r T(r')\,r'^2 \, \df r'
\ee
such that $\lim_{r\to\infty} \tau(r) = 0$, then
\be
\Delta = -2\pi \tau(r\hor) \ .
\ee
The subtle detail here is that $r\hor$ is usually some function of the quantities evaluated at the infinity, $r\hor = r\hor(M,Q,P)$. In principle, we can express the horizon radius $r\hor$ and the charges $Q$ and $P$ with the mass $M$ and the potentials $\Phi\hor$ and $\Psi\hor$ via nonlinear system of equations consisting of the condition for the horizon, $f(r\hor) = 0$, and the expressions for the potentials. However, the generalized Smarr relation obtained in this way will still have a form of an implicit formula for the mass $M$.

\medskip

Here we may make a brief remark about the fall-off conditions in the power-Maxwell model. Suppose we have the purely electric case, the one in which $B_a = 0$. From (\ref{eq:sphBE}) we have $E_r = O(r^{-2/(2s-1)})$ and $\Phi = O(r^{(2s-3)/(2s-1)})$, which gives us restriction $s \in \left< 1/2, 3/2 \right>$ for the convergence of the integral (\ref{eq:QP}). Furthermore, as $r^2 \FF^s = O(r^{-2/(2s-1)})$, we need $s > 1/2$ for the convergence of the integral (\ref{eq:Delta}). In summary, the constraint on the parameter $s$ is that at least $s \in \left< 1/2, 3/2 \right>$.

\bigskip

\subsection{Correction to the Smarr formula via horizon radius}

The authors of \cite{DARG13} have presented a generalization of the Smarr formula for the static, spherically symmetric, purely electric black hole solutions, where the additional term is written explicitly with the horizon radius. They use the fact that $\mathcal{E}_{\mathrm{ex}}(r^2,Q) \equiv M + r(f(r)-1)/2$, refered to as the ``external energy function'', is a homogeneous function of degree $3/2$. Then, by the Euler's theorem it follows \cite{DARG13} that
\be\label{eq:DARGSmarr}
3M = 2(TS + Q\Phi\hor) + r\hor \ .
\ee
We will show how to derive more general relation of this form, which reduces to the equation (\ref{eq:DARGSmarr}) in the purely electric case. Equations (\ref{eq:GTtt}) and (\ref{eq:GTthth}) may be integrated to
\begin{eqnarray}
\frac{1}{2}\,r(f(r) - 1) \Big|_{r\hor}^\infty & = \int_{r\hor}^\infty \LL r^2\,\df r + 4 \int_{r\hor}^\infty (\LL_\FF E_r - \LL_\GG B_r) E_r r^2\,\df r \ , \label{eq:inttt}\\
\frac{1}{4}\,r^2 f'(r) \Big|_{r\hor}^\infty & = \int_{r\hor}^\infty \LL r^2\,\df r - 4 \int_{r\hor}^\infty (\LL_\FF B_r + \LL_\GG E_r) B_r r^2\,\df r \ . \label{eq:intthth}
\end{eqnarray}
In order to assure that the terms are well-defined at the spatial infinity (i.e.~for $r\to\infty$), as part of the boundary conditions we shall assume that the metric function $f$ can be written as
\be
f(r) = 1 - \frac{2M}{r} + O(r^{-(1+\epsilon)})
\ee
for some real $\epsilon > 0$. Then, using the fact that $f(r\hor) = 0$ and that the surface gravity is given by $\kappa = f'(r\hor)/2$, we have
\be
r(f(r) - 1) \Big|_{r\hor}^\infty = -2M + r\hor \qquad \textrm{and} \qquad r^2 f'(r) \Big|_{r\hor}^\infty = 2M - 2\kappa r\hor^2 \ .
\ee
The correction $\Delta$ consists of the following contributions,
\begin{eqnarray}
\Delta & = 2\int_{r\hor}^\infty \LL r^2\,\df r + \nonumber\\
 & + 4 \int_{r\hor}^\infty \Big( (\LL_\FF E_r - \LL_\GG B_r) E_r - (\LL_\FF B_r + \LL_\GG E_r) B_r \Big) r^2\,\df r \ .
\end{eqnarray}
Now, the main point is that we can combine equations (\ref{eq:inttt}) and (\ref{eq:intthth}) so as to express $\Delta$ using only terms on the left hand sides,
\be
\Delta = \frac{1}{2}\,\Big( (1 - \kappa r\hor) r\hor - M \Big) \ .
\ee
Since in the spherically symmetric case $\mathcal{A}\hor = 4\pi r\hor^2$, with this correction the generalized Smarr formula becomes finally
\be\label{eq:rhorSmarr}
3M = \frac{\kappa \mathcal{A}\hor}{4\pi} + 2(\Phi\hor Q\hor + \Psi\hor P\hor) + r\hor \ .
\ee

\bigskip

\subsection{Born-Infeld model}

The correction to the Smarr relation for the static, spherically symmetric black hole solutions in the Born-Infeld model \cite{Born34,BI34,GSP84} was analysed in \cite{YH10,GMK12} and, as we have already remarked above, written as a product of a conjugate pair of thermodynamic variables. The main technical difficulty here is that most of the integrals involve hypergeometric functions. We will just briefly summarize the main results. In the purely electric case equations (\ref{eq:sphBE}) can be easily solved, so that the electric field 1-form is given by
\be
E = \frac{Q}{\sqrt{r^4 + (Q/b)^2}}\,\df r \ ,
\ee
while the corresponding scalar potential on the horizon is
\be
\Phi\hor = -\int_\infty^{r\hor} E_r(r')\,\df r' = \frac{Q}{r\hor}\, {_2 F_1} \left( \frac{1}{4},\frac{1}{2};\frac{5}{4}; -\frac{Q^2}{b^2 r\hor^4} \right) \ .
\ee
The metric function $f$ can be obtained by integrating the equation (\ref{eq:GTtt}) with the appropriate fall-off conditions,
\begin{eqnarray}
f(r) = 1 - \frac{2M}{r} & + \frac{2b^2 r^2}{3} \left( 1 - \sqrt{1 + \frac{Q^2}{b^2 r^4}} \right) + \nonumber\\
 & + \frac{4Q^2}{3r^2} \, {_2 F_1} \left( \frac{1}{4},\frac{1}{2};\frac{5}{4}; -\frac{Q^2}{b^2 r^4} \right) \ .
\end{eqnarray}
Finally, as the Born-Infeld Lagrangian (\ref{eq:BI}) is clearly written in the form 
$$\LL = b^2 \widetilde{\LL}(b^{-2}\FF,b^{-2}\GG) \ ,$$
we may write $\Delta_{\mathrm{BI}} = b \mathcal{C}_{-2}^{\mathrm{BI}}$, with
\be
\mathcal{C}_{-2}^{\mathrm{BI}} = \frac{2b r\hor^3}{3} \left( -1 + \sqrt{1 + \frac{Q^2}{b^2 r\hor^4}} \right) - \frac{Q^2}{3b r\hor}\, {_2 F_1} \left( \frac{1}{4},\frac{1}{2};\frac{5}{4}; -\frac{Q^2}{b^2 r\hor^4} \right) \ .
\ee
Note that the ``Born-Infeld vacuum polarization'' $\mathcal{B}$, as defined in \cite{GMK12}, has the sign opposite to $\mathcal{C}_{-2}^{\mathrm{BI}}$. Unfortunately, the final form of the Smarr formula thus obtained is an ``implicit mess'' and the best one might accomplish for the simplification is the form presented in the equation (\ref{eq:rhorSmarr}).

\bigskip

\subsection{Bardeen's model}

Bardeen's black hole \cite{Bardeen68,ABG00} is the first example of a regular black hole, devoid of the singularity due to the presence of the NLE. This is a spherically symmetric, purely magnetic solution (i.e.~magnetically charged black hole), with
\be
f(r) = 1 - \frac{2Mr^2}{(r^2 + g^2)^{3/2}} \qquad \textrm{and} \qquad B = \frac{g}{r^2}\,\df r \ .
\ee
Here we are using a conventional notation for the magnetic charge, $g = P$. The horizon radius is defined by the condition $f(r\hor) = 0$, which is equivalent to the algebraic condition
\be\label{eq:horBard}
\frac{r\hor}{\sqrt{r\hor^2 + g^2}} = \frac{r\hor^2 + g^2}{2Mr\hor} \ .
\ee
Since the magnetic scalar potential on the horizon is given by
\be
\Psi\hor = \frac{3M}{2g} \left( \frac{r\hor^5}{(r\hor^2 + g^2)^{5/2}} - 1 \right) = \frac{3M}{2g} \left( \frac{(r\hor^2 + g^2)^5}{(2Mr\hor)^5} - 1 \right) \ ,
\ee
it is convenient to introduce an auxiliary quantity
\be\label{eq:x}
x \equiv \left( 1 + \frac{2g\Psi\hor}{3M} \right)^{\!\frac{1}{5}} = \frac{r\hor^2 + g^2}{2Mr\hor} \ ,
\ee
with which we can express the horizon radius,
\be
r\hor = Mx + \sqrt{(Mx)^2 - g^2} \ .
\ee
Then, using (\ref{eq:sphDelta}), we have
\be
\Delta_{\mathrm{Bardeen}} = \frac{M}{2} \left( 1 - \frac{(r\hor^2 + 4g^2)r\hor^3}{(r\hor^2 + g^2)^{5/2}} \right) \ .
\ee
We note in passing that the Bardeen's NLE Lagrangian (\ref{eq:Bard}) can be written in a form $\LL = (M/g) g^{-2} \widetilde{\LL}(g^2\FF)$, so that the same result may be obtained with the method presented in the section 4. Finally, we want to get rid of the horizon radius from the formula for $\Delta_{\mathrm{Bardeen}}$. Using a little bit of algebraic gymnastics with (\ref{eq:horBard}) and (\ref{eq:x}),
\be
\frac{(r\hor^2 + 4g^2)r\hor^3}{(r\hor^2 + g^2)^{5/2}} = \frac{r\hor^3}{(r\hor^2 + g^2)^{3/2}} \left( 1 + \frac{3g^2}{r\hor^2 + g^2} \right) = x^3 \left( 1 + \frac{3g^2}{2Mr\hor x} \right) \ ,
\ee
we can rewrite the $\Delta$ term as
\be
\Delta_{\mathrm{Bardeen}} = \frac{M}{2} \left( 1 - x^3 - \frac{3g^2}{2M}\,\frac{x^2}{Mx + \sqrt{M^2 x^2 - g^2}} \right) \ .
\ee
In result we have managed to reduce the Smarr relation for the Bardeen model to an implicit relation for the mass $M$. The difficulty of expressing the mass $M$ from this form of the Smarr formula is directly related to the difficulty of solving the polynomial equation of some high degree. 

\bigskip

\subsection{Euler-Heisenberg model}

The spherically symmetric solutions for the Euler-Heisenberg NLE model (\ref{eq:EH}) were analysed in \cite{YT00,RWX13,Kruglov17}, however, without any proposed form of the generalized Smarr formula. Suppose that we have a purely electric case, namely $B_r = 0$, thus $\GG = 0$ and $\FF = -2E_r^2$. Then, from the equation (\ref{eq:EHDeltaC}) we immediately have
\be
\Delta_{\mathrm{EH}} = -32\gamma \int_{r\hor}^\infty E_r^4\,r^2\,\df r \ .
\ee
One must take this formula with a grain of salt. Namely, as the Euler-Heisenberg Lagrangian itself is written in the weak-field approximation, this puts the limits on the validity of the last relation in a sense that e.g.~the electric field must be much smaller than the critical field in the black hole exterior (see remarks in \cite{RWX13}). From the equations (\ref{eq:sphBE}) we have 
\be
(1 + 64 \gamma E_r^2) E_r = \frac{Q}{r^2} \ ,
\ee
so that the electric field may be expressed as
\be
E_r = \frac{Q}{r^2} - \frac{\alpha^2}{360 m_e^4}\,\frac{64Q^3}{r^6} + O(\alpha^3)
\ee
Furthermore, by integrating the equation (\ref{eq:GTtt}), we have
\be
f(r) = 1 - \frac{2M}{r} + \frac{Q^2}{r^2} - \frac{\alpha^2}{360 m_e^4}\,\frac{32 Q^4}{5r^6} + O(\alpha^3)
\ee
Suppose that $r_+$ is the larger root of the quadratic equation
\be
1 - \frac{2M}{r} + \frac{Q^2}{r^2} = 0 \ ,
\ee
representing the outer horizon radius of the classical Reissner-Nordstr\"om black hole. Then the solution of the horizon condition $f(r\hor) = 0$ can be written as $r\hor = r_+ + O(\alpha)$. Finally, keeping only the lowest order terms we have the Euler-Heisenberg's correction to the Smarr formula
\be
\Delta_{\mathrm{EH}} = -\frac{\alpha^2}{360 m_e^4}\,\frac{32 Q^4}{5r_+^5} + O(\alpha^3) \ .
\ee

\bigskip

\section{Final remarks}

Using various NLE models we have demonstrated how the novel master formula (\ref{eq:Delta}) for the correction to the Smarr relation can be easily used to deduce and generalize all previously known forms of the NLE Smarr formula. The fact that the generalized Smarr formula remains in its linear form only in some special cases, most notably in the power-Maxwell class of models (covered in the section 5), is probably the reason for the pessimistic attitude of the author of \cite{Rasheed97}. The most intriguing open question is the physical interpretation of the conjugate pair of variables which can be used to write the $\Delta$ term, as discussed in the section 4. Apart from this, a careful analysis   should be done regarding the generalization of the NLE Smarr formula in the presence of the cosmological constant \cite{KMT17,Azreg15}, as well as for the higher dimensional black holes \cite{MP86,GMT99,Kar06,KM06,KNLP05,HEP10,BMMS10,MO11a,MO11b}. It remains to be seen if some of the relations presented in this paper could be tested by the forthcoming astrophysical observations of the black holes.

\ack
This research has been supported by the Croatian Science Foundation under the project No.~8946.

\appendix
\section{Brief compendium of nonlinear electromagnetic models}

For the sake of convenience, we collect in one place the Lagrangians of all the nonlinear electromagnetic models that were explicitly used throughout the paper.

\medskip

\begin{itemize}
\item Born-Infeld Lagrangian (introduced in \cite{Born34,BI34}), an early model of nonlinear electrodynamics, is defined with a real constant $b > 0$ (which corresponds to the strength of the maximal field),
\be\label{eq:BI}
\LL^{\mathrm{(BI)}} = b^2 \left( 1 - \sqrt{1 + \frac{\FF}{2b^2} - \frac{\GG^2}{16b^4}} \right) \ .
\ee
Whenever the condition $\FF \gg (\GG/b)^2$ holds, this Lagrangian can be simplified to the truncated version,
\be\label{eq:tBI}
\LL^{\mathrm{(tBI)}} = b^2 \left( 1 - \sqrt{1 + \frac{\FF}{2b^2}} \right) \ .
\ee

\smallskip

\item Euler-Heisenberg Lagrangian (introduced in \cite{HE36}) is the lowest order quantum correction to the classical Maxwell's Lagrangian, 
\be\label{eq:EH}
\LL^{\mathrm{(EH)}} = -\frac{1}{4}\,\FF + \frac{\alpha^2}{360 m_e^4} \left( 4\FF^2 + 7\GG^2 \right) + O(\alpha^3) \ ,
\ee
where $\alpha$ is the fine-structure constant and $m_e$ is the mass of the electron (for details see e.g.~\cite{RWX13}). We usually use the abbreviation $\gamma = \alpha^2/(360 m_e^4)$.

\medskip

\item Bardeen's model (introduced in \cite{Bardeen68}), is defined with two real parameters $M$ and $g$ (we use the normalization from \cite{ABG00} with $g > 0$),
\be\label{eq:Bard}
\LL^{\mathrm{(Bardeen)}} = \frac{3M}{g^3} \left( \frac{g\sqrt{2\FF}}{2 + g\sqrt{2\FF}} \right)^{\!\frac{5}{2}} \ .
\ee

\smallskip

\item Power-Maxwell Lagrangian (introduced in \cite{HM07,HM08}) is the simplest generalization of the Maxwell's Lagrangian, defined with a pair of real constants $C \ne 0$ and $s \ne 0$,
\be\label{eq:pM}
\LL^{\mathrm{(pM)}} = C\FF^s \ .
\ee
In order to exclude some physically unacceptable solutions, the value of the parameter $s$ is usually restricted to rational numbers with an odd denominator \cite{HM08}.
\end{itemize}

\bigskip

\section{Several useful identities}

The Hodge dual ${*\omega}$ of a $p$-form $\omega$ on a $m$-manifold is defined with
\be\label{eq:Hodge}
(*\omega)_{a_{p+1} \dots a_m} \equiv \frac{1}{p!}\,\omega_{a_1 \dots a_p} \tensor{\epsilon}{^{a_1}^{\dots}^{a_p}_{a_{p+1}}_{\dots}_{a_m}} \ .
\ee
For any two $p$-forms $\alpha$ and $\beta$ we have
\be\label{eq:alphabeta}
\alpha \w {*\beta} = (\alpha \,|\, \beta) \,{*1} \ ,
\ee
with the abbreviation
\be
(\alpha \,|\, \beta) \equiv \frac{1}{p!}\,\alpha_{a_1\dots a_p} \beta^{a_1\dots a_p} \ .
\ee
For any vector $X^a$ and a $p$-form $\alpha_{a_1 \dots a_p}$ we have the Cartan's ``magic formula'',
\be
\Lie_X \alpha = (\df i_X + i_X \df) \alpha
\ee
and a handy ``flipping over the Hodge'',
\be\label{eq:iXHodge}
i_X\,{*\alpha} = *(\alpha \w X) \ ,
\ee
where, by abuse of the notation, the $X$ on the right hand side of the last equation denotes the associated 1-form $X_a = g_{ab} X^b$. The following auxiliary identities are indispensable in algebraic manipulations with the electromagnetic quantities,
\be\label{eq:aux1}
F_{ac} \tensor{F}{_b^c} - {*F}_{ac} \tensor{{*F}}{_b^c} = \frac{1}{2}\,\FF g_{ab} \ ,
\ee
\be\label{eq:aux2}
F_{ac} \tensor{{*F}}{_b^c} = {*F}_{ac} \tensor{F}{_b^c} = \frac{1}{4}\,\GG g_{ab} \ .
\ee

\bigskip

\section*{References}

\bibliographystyle{iopnum}
\bibliography{smarrnle}

\providecommand{\newblock}{}
\begin{thebibliography}{10}
\expandafter\ifx\csname url\endcsname\relax
  \def\url#1{{\tt #1}}\fi
\expandafter\ifx\csname urlprefix\endcsname\relax\def\urlprefix{URL }\fi
\providecommand{\eprint}[2][]{\href{http://arxiv.org/abs/#2}{#2}}

\bibitem{Callen}
Callen H 1985 {\em {Thermodynamics and an introduction to thermostatistics}\/}
  (New York: Wiley) ISBN 978-0471862567

\bibitem{Greiner}
Greiner W 1995 {\em {Thermodynamics and statistical mechanics}\/} (New York:
  Springer-Verlag) ISBN 978-0387942995

\bibitem{Sethna}
Sethna J 2006 {\em {Statistical mechanics: entropy, order parameters, and
  complexity}\/} (Oxford New York: Oxford University Press) ISBN 978-0198566779

\bibitem{Schwabl}
Schwabl F 2006 {\em {Statistical Mechanics}\/} (Berlin New York: Springer) ISBN
  978-3540323433

\bibitem{Lee}
Lee J 2012 {\em Introduction to smooth manifolds\/} (New York London: Springer)
  ISBN 978-1-4419-9981-8

\bibitem{Smarr72}
Smarr L 1973 {\em Phys. Rev. Lett.\/}
  \href{http://dx.doi.org/10.1103/PhysRevLett.30.71}{{\bf {\bf 30}} 71--73}
  [Erratum: Phys. Rev. Lett.30,521(1973)]

\bibitem{Natsuume}
Natsuume M 2015 {\em Lect. Notes Phys.\/}
  \href{http://dx.doi.org/10.1007/978-4-431-55441-7}{{\bf {\bf 903}} 1--294}
  ({\em Preprint} \eprint{1409.3575})

\bibitem{C73}
Carter B 1973 {Black Hole Equilibrium States} {\em {Black Holes}\/} (New York:
  Gordon and Breach) ISBN 9780677156101

\bibitem{C87}
Carter B 1987 {Mathematical Foundations of the Theory of Relativistic Stellar
  and Black Hole Configurations} {\em {Gravitation in Astrophysics}\/}
  (Springer US) ISBN 978-1-4612-9056-8

\bibitem{Heusler97}
Heusler M 1997 {\em Phys. Rev. D\/}
  \href{http://dx.doi.org/10.1103/PhysRevD.56.961}{{\bf {\bf 56}} 961--973}
  \urlprefix\url{https://link.aps.org/doi/10.1103/PhysRevD.56.961}

\bibitem{Heusler}
Heusler M 1996 {\em {Black Hole Uniqueness Theorems}\/} (Cambridge New York:
  Cambridge University Press) ISBN 9780521567350

\bibitem{Komar63}
Komar A 1963 {\em Phys. Rev.\/}
  \href{http://dx.doi.org/10.1103/PhysRev.129.1873}{{\bf {\bf 129}} 1873}

\bibitem{CG17}
Cl{\'e}ment G and Gal'tsov D 2017 {\em Phys. Lett.\/}
  \href{http://dx.doi.org/10.1016/j.physletb.2017.08.041}{{\bf B773} 290--294}
  ({\em Preprint} \eprint{1707.01332})

\bibitem{MZ14}
Ma M~S and Zhao R 2014 {\em Class. Quantum Grav.\/}
  \href{http://dx.doi.org/10.1088/0264-9381/31/24/245014}{{\bf {\bf 31}}
  245014} ({\em Preprint} \eprint{1411.0833})

\bibitem{Wald99}
Wald R~M 2001 {\em Living Rev. Rel.\/}
  \href{http://dx.doi.org/10.12942/lrr-2001-6}{{\bf {\bf 4}} 6} ({\em Preprint}
  \eprint{gr-qc/9912119})

\bibitem{Born34}
Born M 1934 {\em Proc. R. Soc.\/}
  \href{http://dx.doi.org/10.1098/rspa.1934.0010}{{\bf A {\bf 143}} 410--437}

\bibitem{BI34}
Born M and Infeld L 1934 {\em Proc. R. Soc.\/}
  \href{http://dx.doi.org/10.1098/rspa.1934.0059}{{\bf A {\bf 144}} 425--451}

\bibitem{HE36}
Heisenberg W and Euler H 1936 {\em Z. Phys.\/}
  \href{http://dx.doi.org/10.1007/BF01343663}{{\bf {\bf 98}} 714--732} ({\em
  Preprint} \eprint{physics/0605038})

\bibitem{Bardeen68}
Bardeen J~M 1968 {Non-singular General Relativistic Gravitational Collapse}
  {\em {Proceeding of the International Conference GR5}\/} (Tbilisi) p 174

\bibitem{ABG98}
Ay{\'o}n-Beato E and Garc{\'i}a A 1998 {\em Phys. Rev. Lett.\/}
  \href{http://dx.doi.org/10.1103/PhysRevLett.80.5056}{{\bf {\bf 80}}
  5056--5059} ({\em Preprint} \eprint{gr-qc/9911046})

\bibitem{ABG00}
Ay{\'o}n-Beato E and Garc{\'i}a A 2000 {\em Phys. Lett.\/}
  \href{http://dx.doi.org/10.1016/S0370-2693(00)01125-4}{{\bf B493} 149--152}
  ({\em Preprint} \eprint{gr-qc/0009077})

\bibitem{Bronnikov00}
Bronnikov K~A 2001 {\em Phys. Rev. D\/}
  \href{http://dx.doi.org/10.1103/PhysRevD.63.044005}{{\bf {\bf 63}} 044005}
  ({\em Preprint} \eprint{gr-qc/0006014})

\bibitem{BH02}
Burinskii A and Hildebrandt S~R 2002 {\em Phys. Rev. D\/}
  \href{http://dx.doi.org/10.1103/PhysRevD.65.104017}{{\bf {\bf 65}} 104017}
  ({\em Preprint} \eprint{hep-th/0202066})

\bibitem{GSB00}
Garc{\'i}a-Salcedo R and Bret{\'o}n N 2000 {\em Int. J. Mod. Phys.\/}
  \href{http://dx.doi.org/10.1142/S0217751X00002160}{{\bf A {\bf 15}}
  4341--4354} ({\em Preprint} \eprint{gr-qc/0004017})

\bibitem{GSB04}
Garc{\'i}a-Salcedo R and Bret{\'o}n N 2005 {\em Class. Quantum Grav.\/}
  \href{http://dx.doi.org/10.1088/0264-9381/22/22/009}{{\bf {\bf 22}}
  4783--4802} ({\em Preprint} \eprint{gr-qc/0410142})

\bibitem{CGMCL04}
Camara C~S, de~Garcia~Maia M~R, Carvalho J~C and Lima J~A~S 2004 {\em Phys.
  Rev. D\/} \href{http://dx.doi.org/10.1103/PhysRevD.69.123504}{{\bf {\bf 69}}
  123504} ({\em Preprint} \eprint{astro-ph/0402311})

\bibitem{ELNO03}
Elizalde E, Lidsey J~E, Nojiri S and Odintsov S~D 2003 {\em Phys. Lett. B\/}
  \href{http://dx.doi.org/10.1016/j.physletb.2003.08.074}{{\bf {\bf 574}} 1--7}
  ({\em Preprint} \eprint{hep-th/0307177})

\bibitem{LR10}
Labun L and Rafelski J 2010 {\em Phys. Rev. D\/}
  \href{http://dx.doi.org/10.1103/PhysRevD.81.065026}{{\bf {\bf 81}} 065026}
  ({\em Preprint} \eprint{0811.4467})

\bibitem{BLM12}
Bret{\'o}n N, Lazkoz R and Montiel A 2012 {\em JCAP\/}
  \href{http://dx.doi.org/10.1088/1475-7516/2012/10/013}{{\bf {\bf 1210}} 013}
  ({\em Preprint} \eprint{1209.2107})

\bibitem{BR13}
Battesti R and Rizzo C 2013 {\em Rept. Prog. Phys.\/}
  \href{http://dx.doi.org/10.1088/0034-4885/76/1/016401}{{\bf {\bf 76}} 016401}
  ({\em Preprint} \eprint{1211.1933})

\bibitem{FBR16}
Fouch{\'e} M, Battesti R and Rizzo C 2016 {\em Phys. Rev. D\/}
  \href{http://dx.doi.org/10.1103/PhysRevD.93.093020}{{\bf {\bf 93}}(9) 093020}
  \urlprefix\url{https://link.aps.org/doi/10.1103/PhysRevD.93.093020}

\bibitem{MCLP17}
Mosquera~Cuesta H~J, Lambiase G and Pereira J~P 2017 {\em Phys. Rev. D\/}
  \href{http://dx.doi.org/10.1103/PhysRevD.95.025011}{{\bf {\bf 95}} 025011}
  ({\em Preprint} \eprint{1701.00431})

\bibitem{EMY17}
Ellis J, Mavromatos N~E and You T 2017 {\em Phys. Rev. Lett.\/}
  \href{http://dx.doi.org/10.1103/PhysRevLett.118.261802}{{\bf {\bf 118}}
  261802} ({\em Preprint} \eprint{1703.08450})

\bibitem{Rasheed97}
Rasheed D~A 1997  {\em Preprint} \eprint{hep-th/9702087}
  \urlprefix\url{http://arxiv.org/pdf/hep-th/9702087}

\bibitem{Breton04}
Bret{\'o}n N 2005 {\em Gen. Rel. Grav.\/}
  \href{http://dx.doi.org/10.1007/s10714-005-0051-x}{{\bf {\bf 37}} 643--650}
  ({\em Preprint} \eprint{gr-qc/0405116})

\bibitem{GHM09}
Gonz{\'a}lez H~A, Hassa{\"i}ne M and Mart{\'i}nez C 2009 {\em Phys. Rev. D\/}
  \href{http://dx.doi.org/10.1103/PhysRevD.80.104008}{{\bf {\bf 80}} 104008}
  ({\em Preprint} \eprint{0909.1365})

\bibitem{YH10}
Yi-Huan W 2010 {\em Chin. Phys.\/}
  \href{http://dx.doi.org/10.1088/1674-1056/19/9/090404}{{\bf {\bf B19}}
  090404}

\bibitem{GMK12}
Gunasekaran S, Kubiz{\v n}{\'a}k D and Mann R~B 2012 {\em JHEP\/}
  \href{http://dx.doi.org/10.1007/JHEP11(2012)110}{{\bf {\bf 11}} 110} ({\em
  Preprint} \eprint{1208.6251})

\bibitem{DARG13}
Diaz-Alonso J and Rubiera-Garcia D 2013 {\em Gen. Rel. Grav.\/}
  \href{http://dx.doi.org/10.1007/s10714-013-1567-0}{{\bf {\bf 45}} 1901--1950}
  ({\em Preprint} \eprint{1204.2506})

\bibitem{ZG16}
Zhang Y and Gao S 2016  {\em Preprint} \eprint{1610.01237}
  \urlprefix\url{http://arxiv.org/pdf/gr-qc/1610.01237}

\bibitem{FW16}
Fan Z~Y and Wang X 2016 {\em Phys. Rev. D\/}
  \href{http://dx.doi.org/10.1103/PhysRevD.94.124027}{{\bf {\bf 94}} 124027}
  ({\em Preprint} \eprint{1610.02636})

\bibitem{Wald}
Wald R 1984 {\em {General Relativity}\/} (Chicago: University of Chicago Press)
  ISBN 0226870332

\bibitem{Carter70}
Carter B 1970 {\em Commun.Math.Phys.\/}
  \href{http://dx.doi.org/10.1007/BF01647092}{{\bf {\bf 17}} 233--238}

\bibitem{Szabados87}
Szabados L 1987 {\em J.Math.Phys.\/}
  \href{http://dx.doi.org/10.1063/1.527712}{{\bf {\bf 28}} 2688}

\bibitem{MW75}
Michalski H and Wainwright J 1975 {\em Gen. Relativ. Gravit.\/}
  \href{http://dx.doi.org/10.1007/BF00751574}{{\bf {\bf 6}} 289--318}

\bibitem{Tod06}
Tod P 2007 {\em Gen. Relativ. Gravit.\/}
  \href{http://dx.doi.org/10.1007/s10714-006-0363-5}{{\bf {\bf 39}} 111--127}
  ({\em Preprint} \eprint{gr-qc/0611035})

\bibitem{BGS17}
Barja{\v s}i{\'c} I, Gulin L and Smoli{\'c} I 2017 {\em Phys. Rev.\/}
  \href{http://dx.doi.org/10.1103/PhysRevD.95.124037}{{\bf D {\bf 95}} 124037}
  ({\em Preprint} \eprint{1705.00628})

\bibitem{KayWald91}
Kay B~S and Wald R~M 1991 {\em Phys.Rept.\/}
  \href{http://dx.doi.org/10.1016/0370-1573(91)90015-E}{{\bf {\bf 207}}
  49--136}

\bibitem{RW95}
R{\'a}cz I and Wald R~M 1996 {\em Class. Quantum Grav.\/}
  \href{http://dx.doi.org/10.1088/0264-9381/13/3/017}{{\bf {\bf 13}} 539--553}
  ({\em Preprint} \eprint{gr-qc/9507055})

\bibitem{HCC}
Chru{\'s}ciel P~T, Lopes~Costa J and Heusler M 2012 {\em Living Rev. Rel.\/}
  \href{http://dx.doi.org/10.12942/lrr-2012-7}{{\bf {\bf 15}} 7} ({\em
  Preprint} \eprint{1205.6112})
  \urlprefix\url{http://dx.doi.org/10.12942/lrr-2012-7}

\bibitem{ISm12}
Smoli{\'c} I 2012 {\em Class. Quantum Grav.\/}
  \href{http://dx.doi.org/10.1088/0264-9381/29/20/207002}{{\bf {\bf 29}}
  207002} ({\em Preprint} \eprint{1205.1071})

\bibitem{ISm14}
Smoli{\'c} I 2014 {\em Class. Quantum Grav.\/}
  \href{http://dx.doi.org/10.1088/0264-9381/31/23/235002}{{\bf {\bf 31}}
  235002} ({\em Preprint} \eprint{1404.1936})

\bibitem{JG11}
Jaramillo J~L and Gourgoulhon E 2011 {\em Fundam. Theor. Phys.\/}
  \href{http://dx.doi.org/10.1007/978-90-481-3015-3_4}{{\bf {\bf 162}} 87--124}
  ({\em Preprint} \eprint{1001.5429})

\bibitem{BCH73}
Bardeen J~M, Carter B and Hawking S~W 1973 {\em Commun. Math. Phys.\/}
  \href{http://dx.doi.org/10.1007/BF01645742}{{\bf {\bf 31}} 161--170}

\bibitem{HS93}
Heusler M and Straumann N 1993 {\em Class. Quantum Grav.\/}
  \href{http://dx.doi.org/10.1088/0264-9381/10/7/008}{{\bf {\bf 10}}
  1299--1321}

\bibitem{Visser93}
Visser M 1993 {\em Phys. Rev. D\/}
  \href{http://dx.doi.org/10.1103/PhysRevD.48.583}{{\bf {\bf 48}} 583--591}
  ({\em Preprint} \eprint{hep-th/9303029})

\bibitem{Heusler96}
Heusler M 1997 {\em Class. Quantum Grav.\/}
  \href{http://dx.doi.org/10.1088/0264-9381/14/7/001}{{\bf {\bf 14}}
  L129--L134} ({\em Preprint} \eprint{gr-qc/9607001})

\bibitem{Wald90}
Wald R~M 1990 {\em J.Math.Phys.\/}
  \href{http://dx.doi.org/10.1063/1.528839}{{\bf {\bf 31}} 2378--2384}

\bibitem{Torre97}
Torre C~G 1997 {Local cohomology in field theory (with applications to the
  Einstein equations)} {\em {Recent developments in gravitation and
  mathematical physics. Proceedings, 2nd Mexican School, Tlaxcala, Mexico,
  December 1-7, 1996}\/} {\em Preprint} \eprint{hep-th/9706092}

\bibitem{KMT17}
Kubiz{\v n}{\'a}k D, Mann R~B and Teo M 2017 {\em Class. Quantum Grav.\/}
  \href{http://dx.doi.org/10.1088/1361-6382/aa5c69}{{\bf {\bf 34}} 063001}
  ({\em Preprint} \eprint{1608.06147})

\bibitem{deO94}
de~Oliveira H~P 1994 {\em Class. Quantum Grav.\/}
  \href{http://dx.doi.org/10.1088/0264-9381/11/6/012}{{\bf {\bf 11}}
  1469--1482}

\bibitem{ABG99}
Ay{\'o}n-Beato E and Garc{\'i}a A 1999 {\em Phys. Lett.\/}
  \href{http://dx.doi.org/10.1016/S0370-2693(99)01038-2}{{\bf B464} 25} ({\em
  Preprint} \eprint{hep-th/9911174})

\bibitem{YT00}
Yajima H and Tamaki T 2001 {\em Phys. Rev. D\/}
  \href{http://dx.doi.org/10.1103/PhysRevD.63.064007}{{\bf {\bf 63}} 064007}
  ({\em Preprint} \eprint{gr-qc/0005016})

\bibitem{FK03}
Fernando S and Krug D 2003 {\em Gen. Rel. Grav.\/}
  \href{http://dx.doi.org/10.1023/A:1021315214180}{{\bf {\bf 35}} 129--137}
  ({\em Preprint} \eprint{hep-th/0306120})

\bibitem{Dey04}
Dey T~K 2004 {\em Phys. Lett.\/}
  \href{http://dx.doi.org/10.1016/j.physletb.2004.06.047}{{\bf B595} 484--490}
  ({\em Preprint} \eprint{hep-th/0406169})

\bibitem{DARG09}
Diaz-Alonso J and Rubiera-Garcia D 2010 {\em Phys. Rev. D\/}
  \href{http://dx.doi.org/10.1103/PhysRevD.81.064021}{{\bf {\bf 81}} 064021}
  ({\em Preprint} \eprint{0908.3303})

\bibitem{RWX13}
Ruffini R, Wu Y~B and Xue S~S 2013 {\em Phys. Rev. D\/}
  \href{http://dx.doi.org/10.1103/PhysRevD.88.085004}{{\bf {\bf 88}} 085004}
  ({\em Preprint} \eprint{1307.4951})

\bibitem{GSP84}
Garc{\'i}a D~A, Salazar I~H and Pleba{\'n}ski J~F 1984 {\em Nuovo Cimento B
  Serie\/} \href{http://dx.doi.org/10.1007/BF02721649}{{\bf {\bf 84}} 65--90}

\bibitem{Kruglov17}
Kruglov S~I 2017 {\em Mod. Phys. Lett.\/}
  \href{http://dx.doi.org/10.1142/S0217732317500924}{{\bf A32} 1750092} ({\em
  Preprint} \eprint{1705.08745})

\bibitem{Azreg15}
Azreg-A{\"i}nou M 2015 {\em Phys. Rev. D\/}
  \href{http://dx.doi.org/10.1103/PhysRevD.91.064049}{{\bf {\bf 91}} 064049}
  ({\em Preprint} \eprint{1411.2386})

\bibitem{MP86}
Myers R~C and Perry M~J 1986 {\em Annals Phys.\/}
  \href{http://dx.doi.org/10.1016/0003-4916(86)90186-7}{{\bf {\bf 172}} 304}

\bibitem{GMT99}
Gauntlett J~P, Myers R~C and Townsend P~K 1999 {\em Class. Quantum Grav.\/}
  \href{http://dx.doi.org/10.1088/0264-9381/16/1/001}{{\bf {\bf 16}} 1--21}
  ({\em Preprint} \eprint{hep-th/9810204})

\bibitem{Kar06}
Kar S 2006 {\em JHEP\/}
  \href{http://dx.doi.org/10.1088/1126-6708/2006/10/052}{{\bf {\bf 10}} 052}
  ({\em Preprint} \eprint{hep-th/0608018})

\bibitem{KM06}
Kar S and Majumdar S 2006 {\em Phys. Rev.\/}
  \href{http://dx.doi.org/10.1103/PhysRevD.74.066003}{{\bf D74} 066003} ({\em
  Preprint} \eprint{hep-th/0606026})

\bibitem{KNLP05}
Kunz J, Navarro-L{\'e}rida F and Petersen A~K 2005 {\em Phys. Lett.\/}
  \href{http://dx.doi.org/10.1016/j.physletb.2005.03.056}{{\bf B614} 104--112}
  ({\em Preprint} \eprint{gr-qc/0503010})

\bibitem{HEP10}
Hendi S~H and Eslam~Panah B 2010 {\em Phys. Lett.\/}
  \href{http://dx.doi.org/10.1016/j.physletb.2010.01.026}{{\bf B684} 77--84}
  ({\em Preprint} \eprint{1008.0102})

\bibitem{BMMS10}
Banerjee R, Majhi B~R, Modak S~K and Samanta S 2010 {\em Phys. Rev. D\/}
  \href{http://dx.doi.org/10.1103/PhysRevD.82.124002}{{\bf {\bf 82}} 124002}
  ({\em Preprint} \eprint{1007.5204})

\bibitem{MO11a}
Mi{\v s}kovi{\'c} O and Olea R 2011 {\em Phys. Rev.\/}
  \href{http://dx.doi.org/10.1103/PhysRevD.83.024011}{{\bf D83} 024011} ({\em
  Preprint} \eprint{1009.5763})

\bibitem{MO11b}
Miskovic O and Olea R 2011 {\em Phys. Rev.\/}
  \href{http://dx.doi.org/10.1103/PhysRevD.83.064017}{{\bf D83} 064017} ({\em
  Preprint} \eprint{1012.4867})

\bibitem{HM07}
Hassa{\"i}ne M and Mart{\'i}nez C 2007 {\em Phys. Rev. D\/}
  \href{http://dx.doi.org/10.1103/PhysRevD.75.027502}{{\bf {\bf 75}} 027502}
  ({\em Preprint} \eprint{hep-th/0701058})

\bibitem{HM08}
Hassa{\"i}ne M and Mart{\'i}nez C 2008 {\em Class. Quantum Grav.\/}
  \href{http://dx.doi.org/10.1088/0264-9381/25/19/195023}{{\bf {\bf 25}}
  195023} ({\em Preprint} \eprint{0803.2946})

\end{thebibliography}

\end{document}